\documentclass[doublecol]{epl2} 
\usepackage{graphicx}
\usepackage{amsmath,amssymb}

\title{Single-Particle Dynamics in Dense Granular Fluids under 
 Driving}
\author{Matthias Sperl\inst{1} 
\and W. Till Kranz \inst{2,3}
\and Annette Zippelius \inst{2,3}
}                     
\institute{
\inst{1}
Institut f\"ur Materialphysik im Weltraum,
Deutsches Zentrum f\"ur Luft- und Raumfahrt (DLR), 51170 K\"oln, Germany 
\\
\inst{2}
Max-Planck-Institut f\"ur Dynamik und Selbstorganisation,
Bunsenstr. 10, 37073 G\"ottingen, Germany
\\\inst{3}
Georg-August-Universit\"at G\"ottingen, 
Institut f\"ur Theoretische Physik, 
Friedrich-Hund-Platz 1, 37077 G\"ottingen, Germany
}

\date{Received: \today / Revised version: date}
\abstract{ 
We present a mode-coupling theory for the dynamics of a tagged particle in 
a driven granular fluid close to the glass transition. The mean-squared 
displacement is shown to exhibit a plateau indicating structural arrest. 
In contrast to elastic hard-sphere fluids, which are solely controlled by 
volume fraction, the localisation length as well as the critical dynamics 
depend on the degree of dissipation, parametrized by the coefficient of 
normal restitution $\varepsilon$. Hence the resulting glassy structure as 
well as the critical dynamics are nonuniversal with respect to 
$\varepsilon$.
} 
\begin{document}
\maketitle

\section{Introduction}
\label{sec:intro}

Experimental investigations 
\cite{Menon1997a,Goldman2006,Abate2006,Reis2007} as well as simulations 
\cite{Fiege2009,Gholami2011} of dense assemblies of agitated granular 
beads show signatures of a transition from fluid to glassy behavior. 
Motivated by the striking similarity between the measurements for granular 
systems --- necessarily far from equilibrium --- and for equilibrium 
colloidal systems, we have recently generalized the mode-coupling theory 
of the coherent density autocorrelation function $\phi_q(t)$ to driven 
dissipative systems \cite{Kranz2010}. Here, we extend the analysis to the 
single-particle dynamics and compute the incoherent density 
autocorrelation function $\phi^s_q(t)$ as well as the mean squared 
displacement (MSD), which are directly accessible to experiments.

\begin{figure}[htb]
\includegraphics[width=\columnwidth]{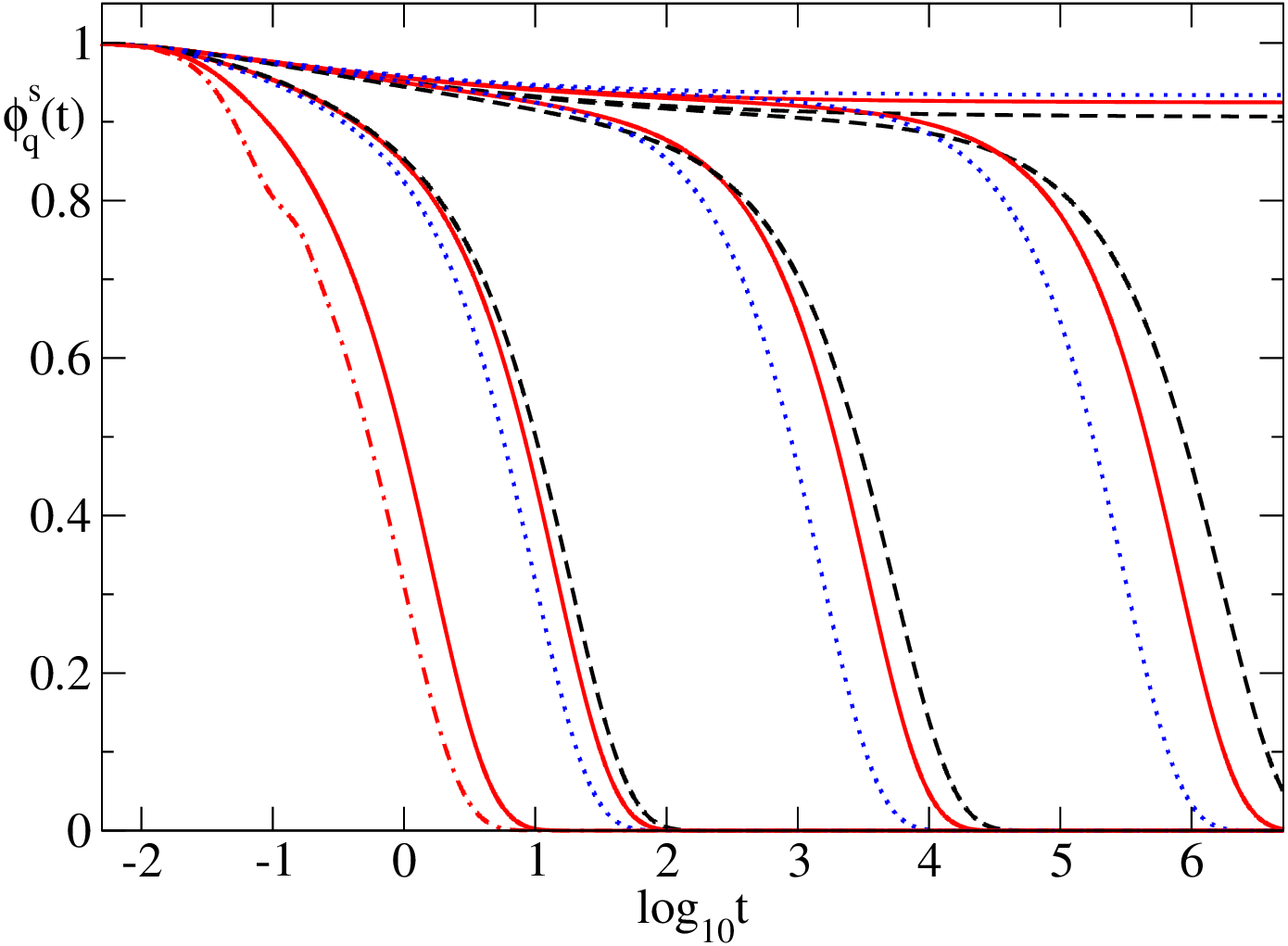}
\caption{\label{fig:phisnu}Incoherent scattering functions $\phi^s_q(t)$ 
for coefficient of restitution $\varepsilon =$ 1.0 (dashed, elastic case), 
0.5 (full curve), and 0.0 (dotted) for packing fractions $\varphi$ from 
right to left: at the glass transition $\varphi^c(\varepsilon)$, and at 
0.999$\varphi^c(\varepsilon)$, 0.99$\varphi^c(\varepsilon)$, and 
0.9$\varphi^c(\varepsilon)$, respectively. Solutions of 
Eq.~(\ref{eq:eoms}) are presented for the wave vector $qd = 4.2$. For 
$\varepsilon = 0.5$, an additional solution is shown for $\varphi = 0.4$ 
and accompanied with a solution where the damping $\nu_q$ is set to zero 
(chain line).
}
\end{figure}

In two dimensions, experimental measurements of the MSD $\delta r^2(t)$ 
are available by direct imaging from air fluidized \cite{Abate2006} and 
mechanically agitated systems \cite{Reis2007}. In three dimensions, the 
MSD is observed by diffusive-wave spectroscopy (DWS) in gravity-driven 
flows of glass beads in ambient air \cite{Menon1997a} and in a 
water-fluidized bed \cite{Goldman2006}. All these studies as well as 
computer simulations \cite{Gholami2011} in two-dimensional systems reveal 
the development of a plateau in the MSD and relate their findings to 
glassy dynamics. With analogous results from colloidal dispersions in 
mind, this has been interpreted as the signature of a granular glass 
transition. Measurements of additional observables support this 
interpretation. Namely, the incoherent scattering function has been found 
to develop a two-step relaxation at high densities both in experiments 
\cite{Reis2007} and in simulations \cite{Gholami2011}. Moreover, plateaus 
extending over increasingly lager windows in time imply a strong decrease 
of the diffusion coefficient $D = \lim_{t\to\infty}\delta r^2(t)/6t$ 
\cite{Abate2006,Fiege2009,Gholami2011} accompanied by a strong increase of 
relaxation times, $\tau$, \cite{Reis2007,Gholami2011} upon approaching the 
granular glass transition density.

For systems in thermal equilibrium, mode-coupling theory (MCT) has become 
an established tool for the investigation of glassy dynamics, it describes 
many experimental features and has the potential for non-trivial 
predictions \cite{Goetze2009}. Applied to the case of colloidal 
suspensions, it is found that MCT is quantitatively accurate to about 20\% 
in the density. For the mean-squared displacement, MCT describes the 
measured data for the entire regime available which is over eight orders 
of magnitude in time \cite{Megen1998,Sperl2005}.

The granular mode coupling theory for homogeneously driven systems, 
outlined in Ref.~\cite{Kranz2010}, predicts (1) the existence of a glass 
transition, i.e., $\lim_{t\rightarrow\infty}\phi_q(t) = f_q >0$, (2) an 
increase of the packing fraction at the transition, 
$\varphi^c(\varepsilon)$, with increased dissipation quantified by the 
normal coefficient of restitution $\varepsilon$, and (3) changes to the 
dynamical exponents with dissipation.

While the glass transition within MCT is found to be a singularity in the 
coherent functions $\phi_q(t)$, incoherent scattering functions 
$\phi^s_q(t)$ and in particular the MSD, $\delta r^2(t)$, have been 
measured in the experiments and simulations discussed above. In order to 
compare to such data, the MCT for incoherent functions is derived for 
granular systems in the following.

\section{Model}\label{sec:model}

We consider a granular fluid consisting of $N$ hard spheres of mass $m=1$ 
and diameter $d$ in a volume $V$. Particle positions and velocities are 
denoted by $\{\vec r_i,\vec v_i\}$ and we will consider the thermodynamic 
limit such that the density $n=N/V$ is finite. Energy dissipation in 
binary collisions is modeled by incomplete normal restitution, quantified 
by a constant coefficient of restitution $\varepsilon$. To achieve a 
stationary state the system is driven randomly and homogeneously: All 
particles are kicked stochastically at random time intervals.  This allows 
the system to relax to a stationary state with finite temperature, $T$, 
defined as the average kinetic energy of the particles.

It is important to note that the system is not in equilibrium so that the 
$N$-particle distribution function, $w(\Gamma)$, is in general unknown.  
We assume \cite{Brilliantov2007} that positions and velocities are 
uncorrelated, $w(\Gamma) = w_r(\{\vec r_i\})w_v(\{\vec v_i\})$, and that 
the velocity distribution factorizes into a product of one-particle 
distributions, $w_v(\{\vec v_i\}) = \prod_iw_1(\vec v_i)$. The precise 
from of $w_1(\vec v)$ is not needed, it only has to satisfy 
$\frac{1}{N}\sum_i {\vec v_i^2} = 3T$, finite.  Furthermore, the system is 
assumed to be isotropic and homogeneous except for the excluded volume 
interaction: $w_r(\{\vec r_i\}) = \prod_{i<j}\theta(r_{ij}-d)$.

\section{Granular Mode-Coupling Theory}
\label{sec:gran-mode-coupl}

The dynamics can be formulated in terms of a Pseudo-Liouville operator 
\cite{Huthmann1997}, so that techniques for the derivation of MCT for 
energy conserving systems with Newtonian dynamics in \cite{Chong2001} can 
be used with appropriate changes \cite{Kranz2010}. The central quantity 
to describe the dynamics of a tagged particle is the incoherent 
intermediate scattering function
\begin{equation}
 \phi^s_q(t):=\left[
\int d\Gamma w(\Gamma) 
\frac{1}{N}\sum_i\exp(i\vec q\cdot(\vec r_i(t)-\vec r_i(0))\right],
\nonumber
\end{equation}
where $[\dots]$ denotes the average over the random driving.

Using appropriate Mori-projectors, one finds that $\phi^s_q(t)$ obeys
the following equation of motion
\begin{subequations}\label{eq:eom}
\begin{equation}\label{eq:eoms}\begin{array}{l}
\partial_t^2 \phi^s_q(t) + \nu^s_q \partial_t 
\phi^s_q(t) + \Omega_{s\,q}^2 \phi_q^s(t)\\ 
\qquad+ \Omega_{s\,q}^2 \int_0^t\,dt'
m^s_q(t-t')\partial_{t'}\phi_q^s(t')=0,\end{array}
\end{equation}
formally identical to the equation of motion for an energy conserving 
system. Here the friction, $\nu^s_q $, is given by the same expression as 
for the coherent part:
\begin{equation}\label{eq:eomnu}
\nu^s_q =  \nu_E
\frac{1+\varepsilon}{2}[1 - j_0(qd) +  2j_2(qd)]
\end{equation}
with $j_i$ the spherical Bessel functions and with the classical 
(equilibrium) Enskog collision rate $\nu_E = 4\sqrt{\pi}nd^2g_dv_0$ 
\cite{Boon1980} where $g_d$ is the contact value of the pair distribution 
function $g(r)$ and $v_0 = \sqrt{T/m}$ is the thermal velocity. Following 
Ref.~\cite{Kranz2010}, we applied the mode-coupling approximation to the 
memory kernel $m^s_q(t)$,
\begin{equation}\label{eq:mems}
  m^s_q(t) \simeq \frac{1+\varepsilon}{2} \frac{n}{q^2}
	\int\, \frac{d^3k}{(2\pi)^3}\;(\hat{\bm{q}}\cdot\bm{k})^2
	S_k c_{k}^2\phi_k(t)\phi^s_{|\bm{q}-\bm{k}|}(t).
\end{equation}
\end{subequations}
Details of the derivation from the microscopic dynamics will be published 
elsewhere. Here the $\phi_q(t)$ are given by the solutions of the coherent 
MCT equations \cite{Kranz2010}, $S_k$ denotes the static structure factor 
and $c_{k}$ the direct correlation function. In contrast to the coherent 
case, the frequency $\Omega_{s\,q}^2 = q^2T$ does not carry any dependence 
on the coefficient of restitution $\varepsilon$. The equations of motion, 
Eq.~(\ref{eq:eom}), are solved with the initial conditions $\phi^s_q(0) = 
1, \partial_t\phi^s_q(0) = 0$. The numerical algorithms for solving the 
equations of motion for the coherent and the incoherent intermediate 
scattering function as well as for the MSD have been introduced previously 
\cite{Franosch1997,Sperl2005,Kranz2010}.

\section{Intermediate Scattering Functions}
\label{sec:ISF}

\begin{figure}[htb]
\includegraphics[width=\columnwidth]{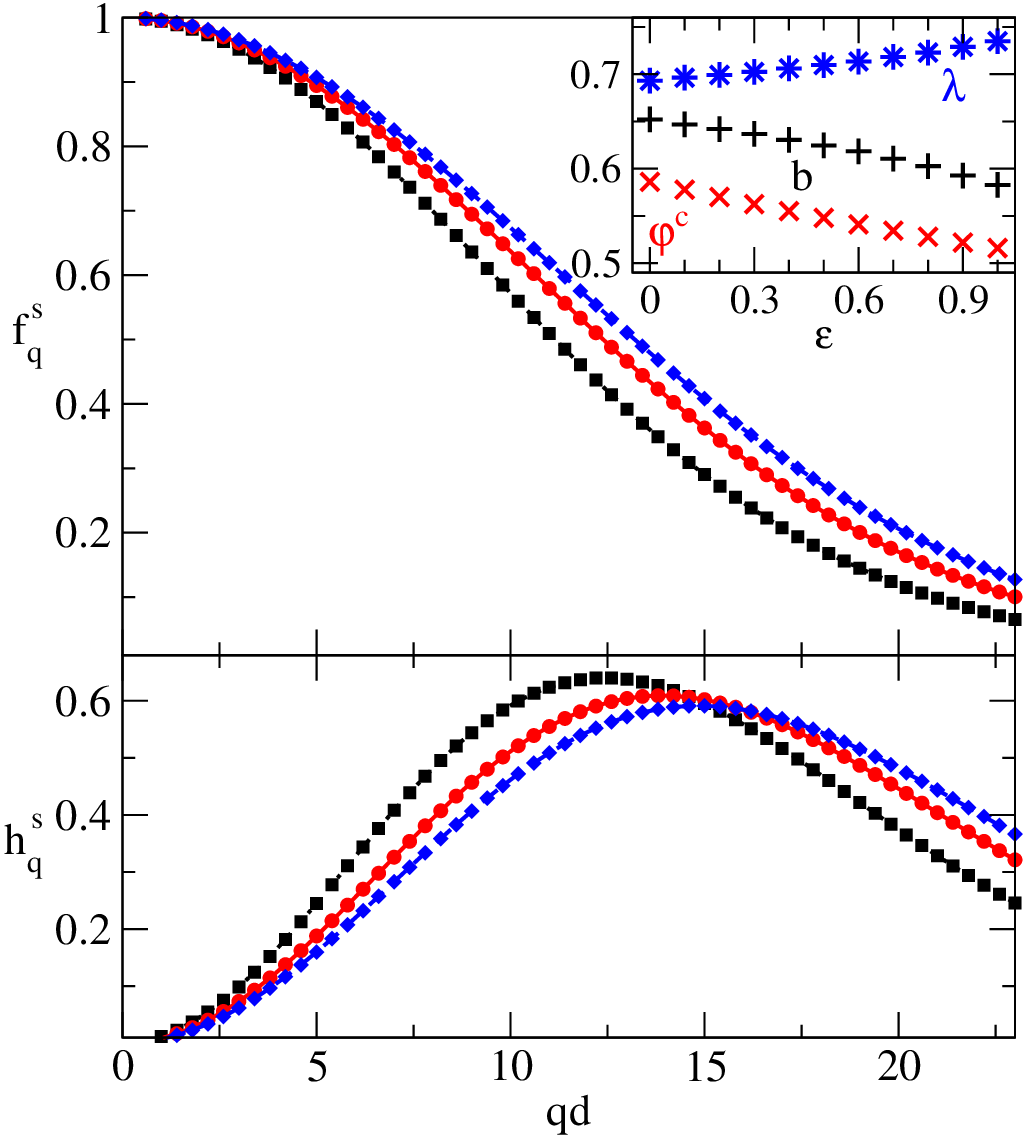}
\caption{\label{fig:fseps}Glass form factors $f^s_q$ (upper panel) and 
critical amplitudes $h^s_q$ (lower panel), cf. Eq.~(\ref{eq:asy}), for the 
incoherent scattering functions for coefficient of restitution 
$\varepsilon =$ 1.0 (squares), 0.5 (circles), and 0.0 (diamonds). The 
inset shows the exponent parameter $\lambda$ (*), the von-Schweidler 
exponent $b$ (+), and the packing fraction at the transition $\varphi^c$ 
($\times$) depending on the coefficient of restitution $\varepsilon$.
}
\end{figure}

Fig.~\ref{fig:phisnu} shows the solutions for the wave vector $qd = 4.2$ 
which is at about half of the principal peak in the static structure 
factor $S_q$. For increasing packing fraction $\varphi = \pi nd^3/6$, the 
scattering functions develop a plateau for $0\leq\varepsilon\leq 1$. For 
smaller $\varepsilon$, the critical plateau value $f^s_q = \lim_{t \to 
\infty} \phi_q^s(t)$ at the transition point increases. It is seen in 
Fig.~\ref{fig:fseps} that this increase with $\varepsilon$ applies to all 
wave vectors for $f_q^s$. The increase results in a growing half width of 
$f^s_q$ as a function of wavenumber, which is related to the inverse 
localization length of a tagged particle.

One important result of MCT is the existence of two timescales, which both 
diverge with the distance from the critical point, $\sigma=(\varphi_c - 
\varphi)/\varphi_c$. The first one, $t_{\sigma}\propto \sigma^{-1/(2a)}$ 
rules the dynamics near the plateau, whereas $\tau \propto 
\sigma^{-\gamma}, \gamma=1/(2a) + 1/(2b)$ determines the asymptotic time 
dependence, i.e., the $\alpha-$relaxation and the critical behavior of the 
diffusion coefficient. As for the elastic case, both exponents $a$ and $b$ 
are expressed by a single exponent parameter $\lambda$, which varies with 
$\varepsilon$ as shown in the inset of Fig.~\ref{fig:fseps}.

The plateau values, or glass form factors, $f^s_q$ form the basis for the 
asymptotic expansion of the MCT equations of motion around the plateau 
\cite{Fuchs1998}. Close to the transition point, the correlation function in 
the vicinity of the plateau can be expanded in leading order as
\begin{subequations}\label{eq:asy}
\begin{equation}\label{eq:asys}
\phi^s_q(t;\sigma) = f^s_q + h^s_q G_\sigma(t)
\end{equation}
which defines the critical amplitudes $h^s_q$ shown in the lower panel of 
Fig.~\ref{fig:fseps}. While $f^s_q$ and $h^s_q$ are fixed by the details 
of the equations of motion at the transition point, $\sigma=0$, the 
scaling function $G_\sigma(t)$ depends only on the time $t$ and the 
distance $\sigma$ from the transition point. The scaling 
law~(\ref{eq:asys}) reveals a factorization property unique to glassy 
dynamics, where the complex dynamics can be separated into a 
time-dependent and a wavenumber-dependent part.  Directly at the 
transition point, Eq.~(\ref{eq:asys}) reduces to the critical law
\begin{equation}\label{eq:asycrit}
\phi^s_q(t;\sigma) = f^s_q + h^s_q (t/t_0)^{-a}\,,
\end{equation}
with a microscopic time scale $t_0$ and the exponent $a$ which is shown in 
the inset of Fig.~2 of \cite{Kranz2010}. In the fluid state, close to the 
transition, the above power law describes the approach to the plateau 
value. In the fluid state below the transition, $G_\sigma(t)$ describes 
the decay from the plateau to zero and in this regime is known as the 
von-Schweidler law
\begin{equation}\label{eq:asyvS}
\phi^s_q(t;\sigma) = f^s_q - h^s_q (t/\tau)^b\,.
\end{equation}
\end{subequations}

\begin{figure}[htb]
\includegraphics[width=\columnwidth]{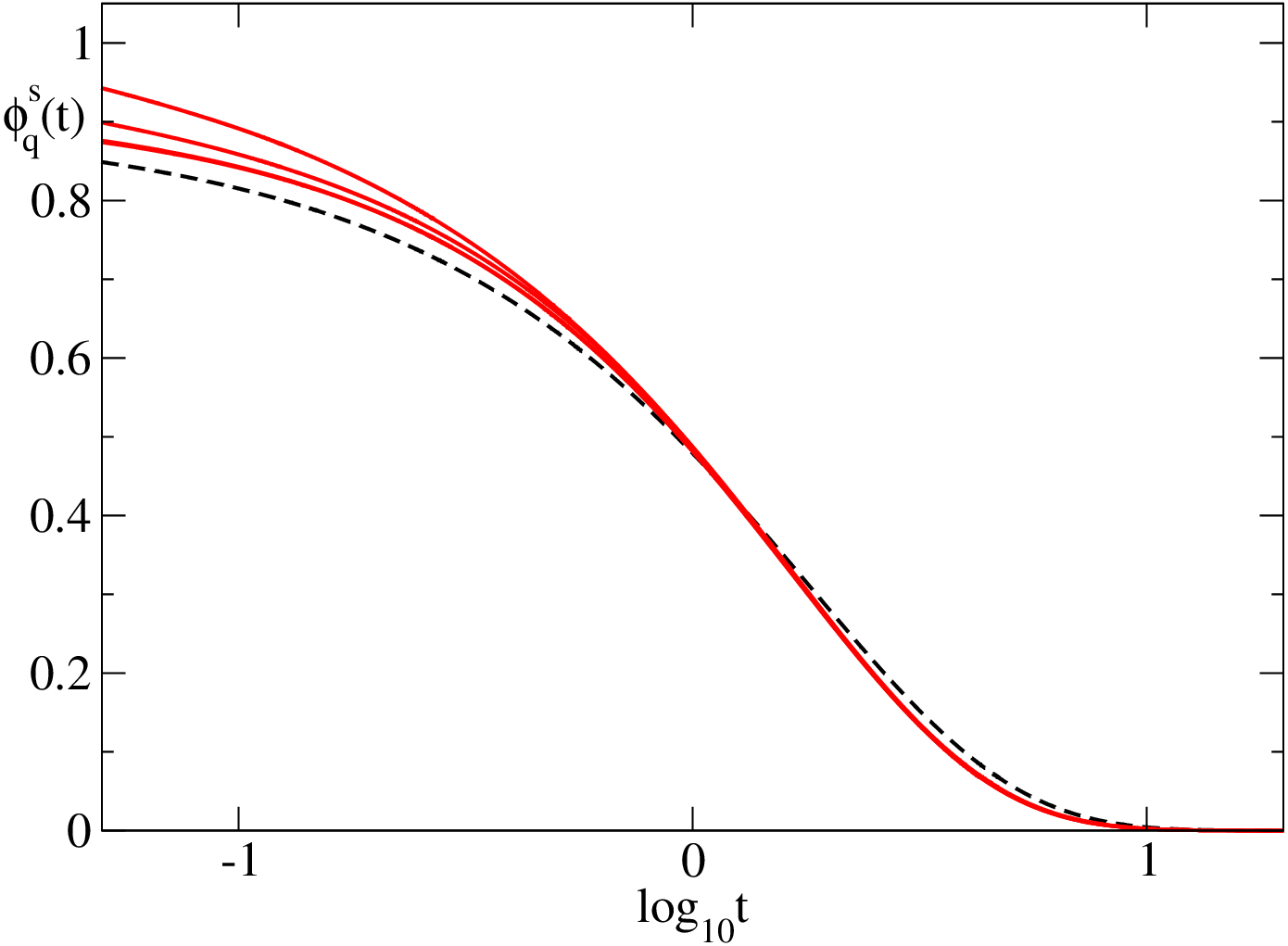}
\caption{\label{fig:alpha}$\alpha$-scaling of the correlators of Fig. 
\ref{fig:phisnu} for coefficient of restitution $\varepsilon = 0.5$
(full curves). Curves are matched at $\phi^s_q(t) = 0.4$. For
comparison, the last fluid curve for $\varepsilon = 1.0$ is shown
dashed.
}
\end{figure}

For asymptotically long times, $t\gg t_{\sigma}$, the incoherent scattering 
functions obey the socalled $\alpha$-scaling
\begin{equation}
\label{eq:asymt}
\phi^s_q(t;\sigma)=\tilde{\phi}^s_q(t/\tau(\sigma))
\end{equation}
which connects to the von-Schweidler law for $x=t/\tau\leq1$, whereas for 
large $x$, the decay is close to a Kohlrausch law \cite{Fuchs1994}, and 
crossing over to an exponential for the largest $x$. The scaling suggested 
by Eq.~(\ref{eq:asymt}) is applied to the results for $\varepsilon = 0.5$ 
from Fig.~\ref{fig:phisnu} and displayed in Fig.~\ref{fig:alpha}. Time 
scales $\tau$ are determined where the correlation functions cross the 
value $\tilde\phi^s_q(\tau) = 0.4$, and the curves are scaled on top of 
the first correlator accordingly. When getting closer to the transition, 
the correlation functions follow a master curve for progressively longer 
times. In contrast, an equally extended correlation function for the 
elastic case, i.e., a different value for $\varepsilon$, clearly violates 
that scaling. With the form factors, $f^s_q$, and amplitudes, $h^s_q$, as 
well as the exponents $b$, all being nontrivial functions of $\varepsilon$ 
no such scaling is expected.

\begin{figure}[htb]
\includegraphics[width=\columnwidth]{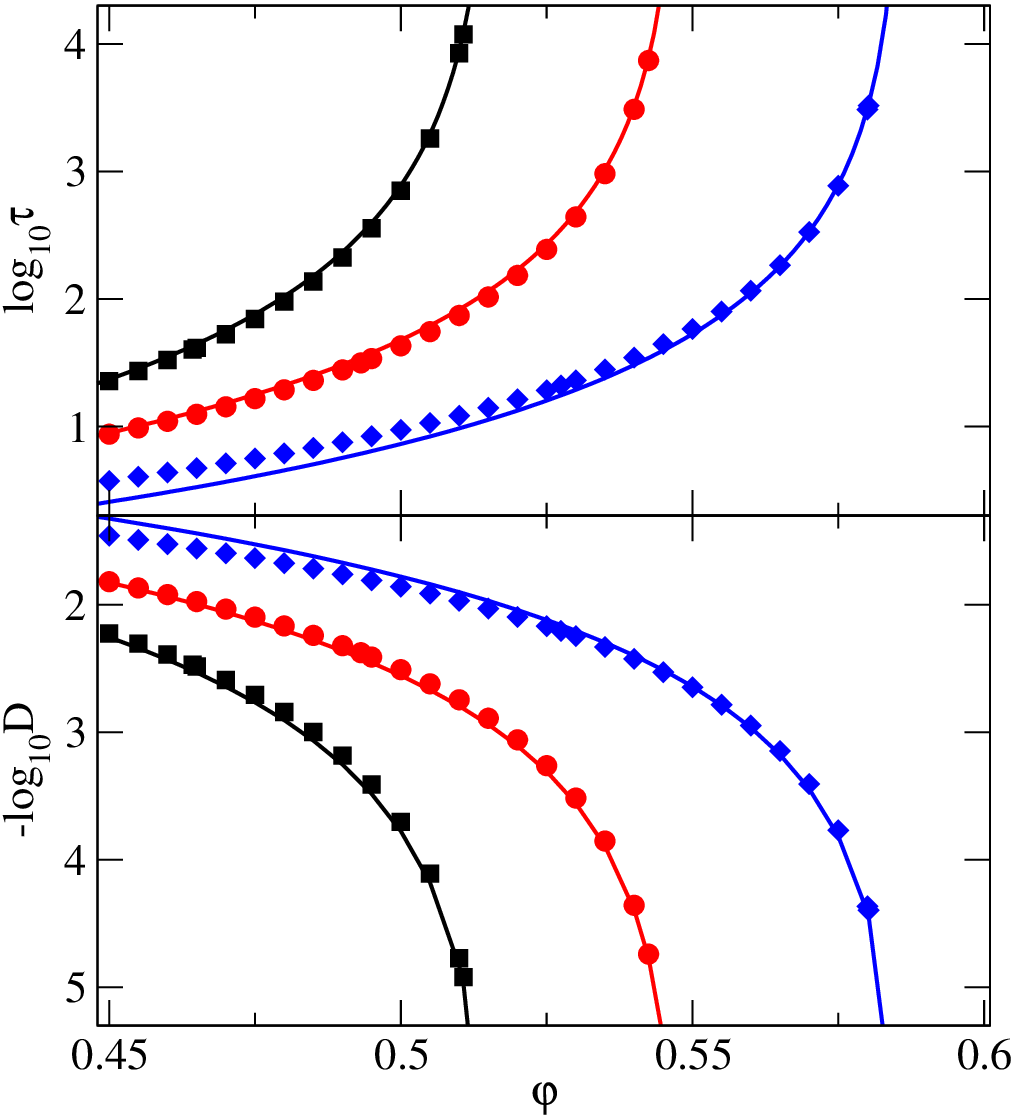}
\caption{\label{fig:Dtau} Dependence on the packing fraction, $\varphi$, 
of the time scale $\tau$ where the incoherent scattering function 
$\phi^s_q(\tau)=0.1$ (upper panel), and diffusion coefficients $D$ (lower 
panel) derived from the numerical solutions of the equations of motion for 
coefficient of restitution $\varepsilon = 1.0$ (squares), 0.5 (circles), 
and 0.0 (diamonds). The full lines show the corresponding asymptotic laws 
$\tau \propto [\varphi^c(\varepsilon) - \varphi(\varepsilon)]^{-\gamma}$ 
and $D\propto [\varphi^c(\varepsilon) - \varphi(\varepsilon)]^\gamma$. For 
$\varepsilon = 1.0$, 0.5, and 0.0, the values for $\gamma$ are 2.46239, 
2.34921, and 2.28282, respectively.
}
\end{figure}

The divergence of the time scale, $\tau$, determines the vanishing of the 
diffusion constant $D$. Both singularities are governed by the asymptotic 
law $D\propto \tau^{-1}\propto \sigma^\gamma$ which is shown by full lines 
in Fig.~\ref{fig:Dtau}.  Individual symbols in the same plots show the 
actual values retrieved from the numerical solution of the equations of 
motion. The upper panel demonstrates that for sufficiently large time 
scales the asymptotic law describes the numerical values satisfactorily. 
On the contrary, for distances around 10\% from the critical point and 
larger, one cannot expect the asymptotic law to hold --- and the early 
part of the divergence might even suggest a power law with a different 
exponent.

The evolution of the diffusion coefficients in the lower panel of 
Fig.~\ref{fig:Dtau} suggests a quite similar behavior as for $\tau$. Again 
the asymptotic law works well very close to the transition while for 
larger distances of 10\% and more, the numerical solution diverges 
markedly slower than the asymptotic law would suggest.

The structure of the Equations (\ref{eq:asy},\ref{eq:asymt}) is the same 
as for the incoherent functions of the elastic hard-sphere system 
\cite{Fuchs1998}, including the factorization property for the dynamics 
around the plateau. However, the exponent parameter $\lambda$ as well as 
glass form factor ($f^s_q$, cf. Fig.~\ref{fig:fseps}) and the critical 
amplitude ($h^s_q$, cf. Fig~\ref{fig:fseps}) do depend on the coefficient 
of restitution. Hence we conclude that the dynamics is not universal with 
respect to dissipation.

\begin{figure}[htb]
\includegraphics[width=\columnwidth]{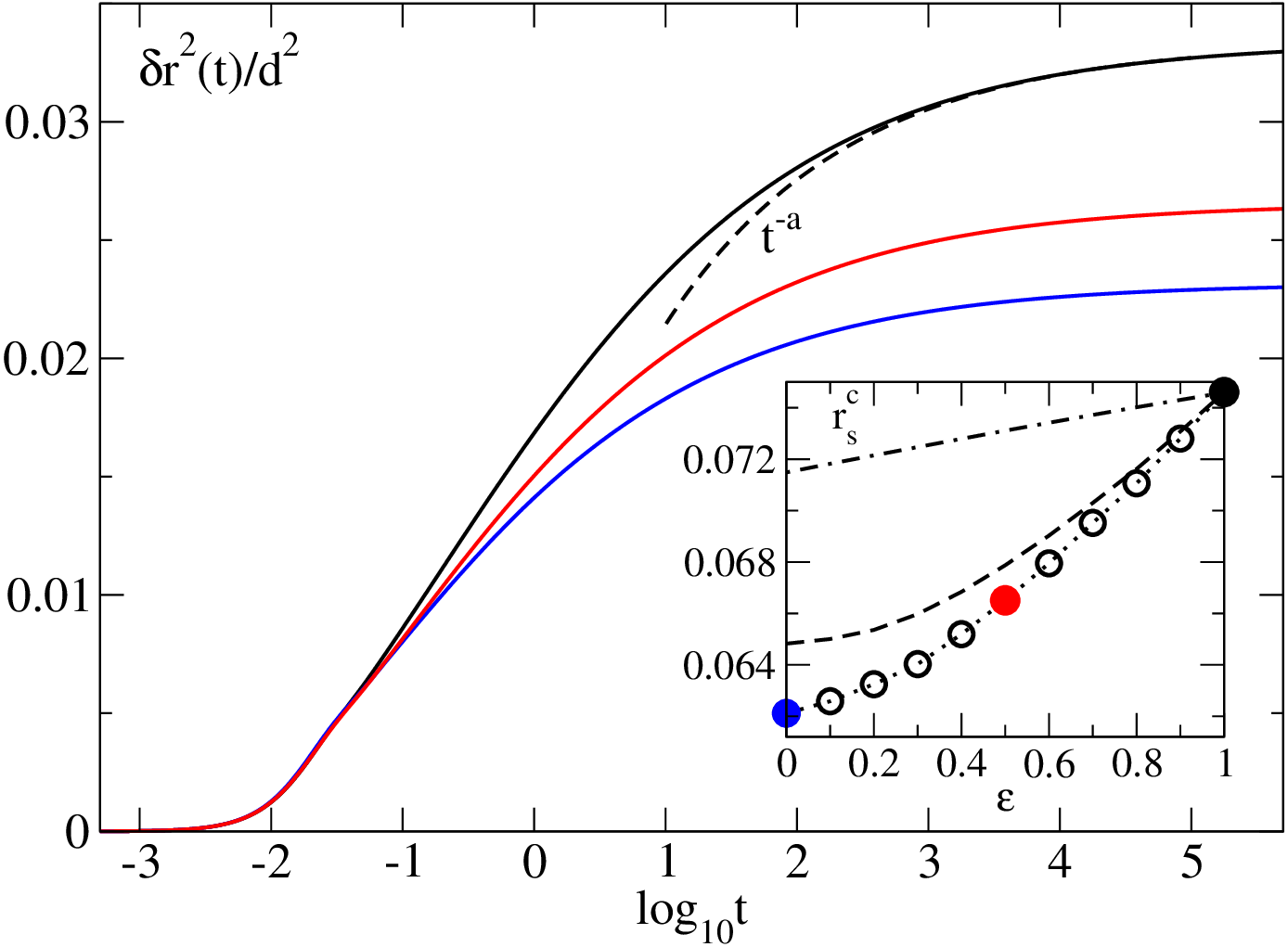}
\caption{\label{fig:MSDcritloc}Mean-squared displacement $\delta r^2(t)$ 
at the glass-transition packing fraction $\varphi^c(\varepsilon)$ for 
coefficient of restitution $\varepsilon = 1.0$, 0.5, and 0.0 decreasing 
from top to bottom at long times. Full curves show the numerical solutions 
of Eq.~(\ref{eq:eomMSD}). The dashed curve displays the critical law in 
Eq.~(\ref{eq:asyMSDcrit}) with $t_0 = 0.035$ for the case $\varepsilon = 
1.0$. The inset shows the localization length, cf. Eq.~(\ref{eq:loc}), as 
a function of $\varepsilon$ with $\varepsilon = 1.0$, 0.5, and 0.0 shown 
as filled circles. The chain curve shows the elastic limit 
$r_c(\varepsilon=1)$ scaled by the evolution of the mean particle 
separation, $\ell_0 \propto 
1/\sqrt[3]{\varphi^c(\varepsilon)/\varphi^c(1)}$. The dashed curve 
indicates the calculated localization length with the latter dependence 
scaled out.
}
\end{figure}

\section{Mean-Squared Displacement}\label{sec:MSD}

Experiments as well as simulations focus on the MSD which is defined by
\begin{equation}
\delta r^2(t):=\left[
\int d\Gamma w(\Gamma) 
\frac{1}{N}\sum_i(\vec r_i(t)-\vec r_i(0))^2\right]\nonumber
\end{equation}	
and can also be obtained from the expansion of $\phi^s_q(t) = 1 - 
q^2\delta r^2(t) + \mathcal O(q^4)$ for small wavenumbers.  The equation 
of motion for the MSD in the granular case reads
\begin{subequations}\label{eq:eomMSD}
\begin{equation}\label{eq:gMSD}\begin{array}{l}
\partial_t\delta r^2(t) 
+ \frac{1+\varepsilon}{2} \nu_E \delta r^2(t) \\\qquad
+ v_0^2\int_0^t\,dt'm^{(0)}(t-t')\delta r^2(t') 
= 6 v_0^2 t\end{array}
\end{equation}
and the memory kernel within MCT is given by
$m^{(0)}(t) = \lim_{q\to0}q^2m^s_q(t)$, and reads
\begin{equation}\label{eq:gMSDkernel}
m^{(0)}(t) =  \frac{1+\varepsilon}{2} 
	\frac{n}{6\pi^2}\int_0^{\infty}\,dk\;k^4 S_k 
c_k^2\phi_k(t)\phi^s_k(t)\,.
\end{equation}
\end{subequations}

\begin{figure}[htb]
\includegraphics[width=\columnwidth]{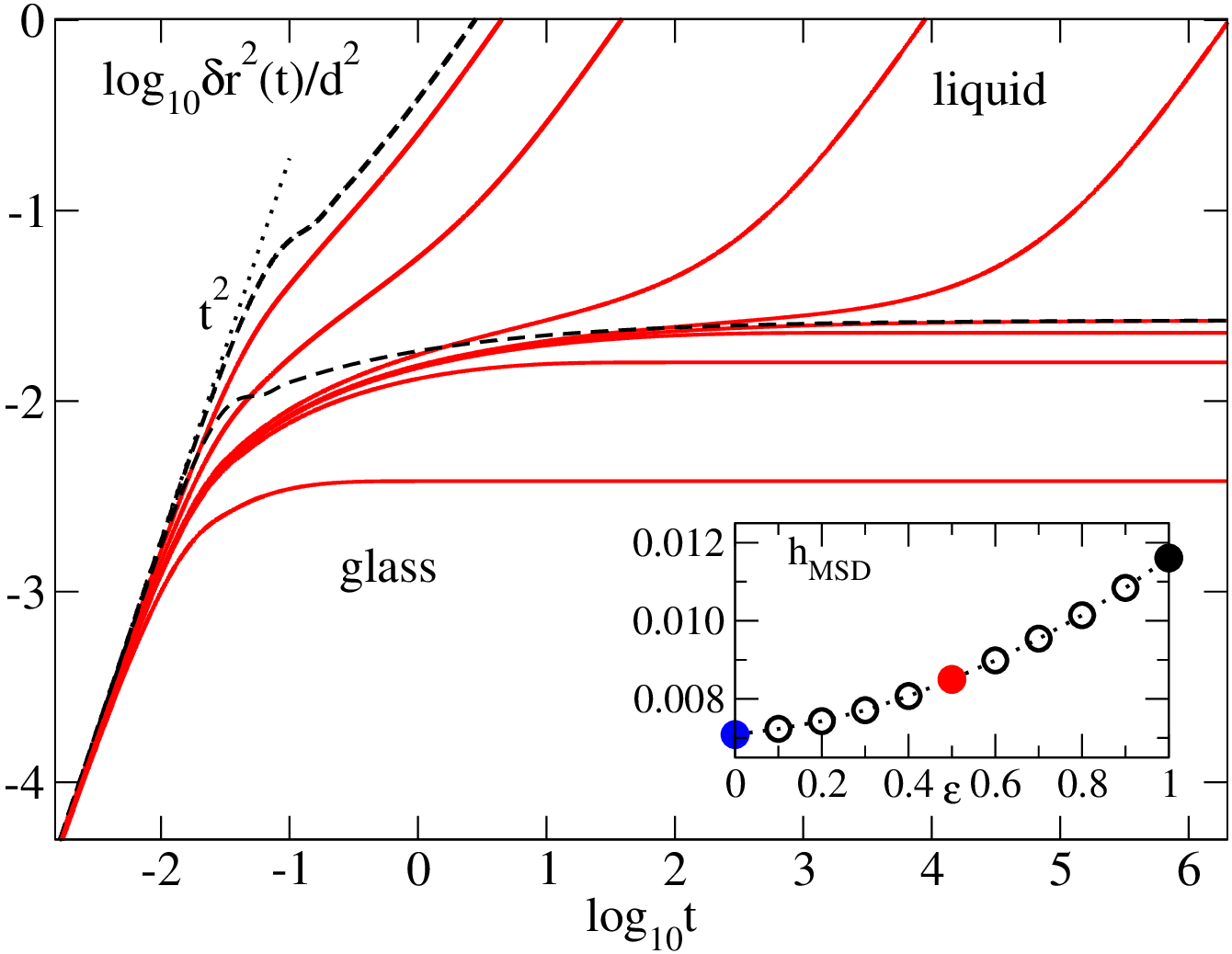}
\caption{\label{fig:MSDnueps05}Mean-squared displacement $\delta r^2(t)$ 
for coefficient of restitution $\varepsilon = 0.5$ and packing fractions 
$\varphi = \varphi^c$, 1.001$\varphi^c$, 1.01$\varphi^c$, 1.1$\varphi^c$, 
on the arrested side --- 0.999$\varphi^c$, 0.99$\varphi^c$, 
0.9$\varphi^c$, and 0.4, respectively, on the diffusive side. The terms 
liquid and glass indicate the diffusive and arrested regimes, 
respectively. Dashed curves show the results for the critical point and 
$\varphi = 0.4$ with the choice of $\nu_q = \nu^s_q = \nu_E=0$. The 
initial ballistic law $v_0^2t^2$ is shown by a dotted straight line. The 
inset exhibits the critical amplitude $h_\text{MSD}$ as a function of 
$\varepsilon$.
}
\end{figure}

The behavior of the MSD at the respective critical points for several 
$\varepsilon$ is shown in Fig.~\ref{fig:MSDcritloc}.  The effective Enskog 
damping coefficient $\nu_E(1+\varepsilon)/2$ decreases for smaller 
$\varepsilon$, and this lesser damping causes the MSD for $\varepsilon=1$ 
to be slightly larger than that for $\varepsilon=0.5$ and $\varepsilon=1$ 
in the time window $-2\leq\log_{10} t\leq-1$ after the ballistic regime. 
For macroscopic times the solutions reverse their order and finally reach 
their long-time limits $6r^{2}_c$ with the localization length being 
defined by
\begin{subequations}\label{eq:asyMSD}
\begin{equation}\label{eq:loc}
r^{2}_c = \lim_{t\rightarrow\infty}\delta r^2(t)/6 = 1/ 
\lim_{t\rightarrow\infty}m^{(0)}(t)\,.
\end{equation}
Similarly to the glass-form factors, for the MSD an asymptotic expansion 
can be performed with the result similar to Eq.~(\ref{eq:asy})
\begin{equation}\label{eq:asyMSDG}
\delta r^2(t)/6 = r^{2}_c - h_\text{MSD} G_\sigma(t)\,.
\end{equation}
Together with the critical,
\begin{equation}\label{eq:asyMSDcrit}
\delta r^2(t)/6 = r^{2}_c - h_\text{MSD} (t/t_0)^{-a}\,,
\end{equation}
and the von-Schweidler law,
\begin{equation}\label{eq:asyMSDvS}
\delta r^2(t)/6 = r^{2}_c + h_\text{MSD} (t/\tau)^b\,,
\end{equation}
\end{subequations}
the analysis of the elastic hard-sphere systems in \cite{Fuchs1998} 
carries over to the dissipative case. The evolution of $r_c$ and 
$h_\text{MSD}$ with $\varepsilon$ are shown in the insets of 
Figs.~\ref{fig:MSDcritloc} and~\ref{fig:MSDnueps05}. In addition, the 
limited applicability of the asymptotic critical law due to large 
corrections to scaling applies to elastic as well as dissipative hard 
spheres \cite{Sperl2005}.

It is seen in the inset of Fig.~\ref{fig:MSDcritloc} that $r_c$ decreases 
with $\varepsilon$ which is corresponding to a higher packing fraction at 
the glass transition (cf. inset of Fig.~\ref{fig:fseps}). While the 
overall decrease of $r_c$ from $\varepsilon = 1$ to $\varepsilon = 0$ is 
more than 16\%, the trivial contribution from the increase in density can 
explain only 4\% of the decrease: A length scale can be defined by the 
inverse cubic root of the packing fraction which gives a scaling factor 
$\sqrt[3]{\varphi^c(\varepsilon)/\varphi^c(\varepsilon=1)}$. When the 
calculated results are scaled with this factor, the dashed curve in the 
inset of Fig.~\ref{fig:MSDcritloc} is found. Hence, the major part of the 
decrease of the localization length is a more involved prediction than 
just a simple density scaling argument.

\section{Short-Time Dynamics}\label{sec:short}

It is known that the damping $\nu_q$ together with the early part of the 
memory kernel can overestimate the total damping considerably, which is an 
issue that cannot be fully resolved \cite{Goetze2009}. It is possible to 
estimate the order of magnitude of that effect without introducing 
mathematical inconsistencies by setting all the damping terms, $\nu_q$, 
$\nu^s_q$, $\nu_E$ to zero and solve the equations. The result is shown as 
the chain curve for $\phi^s_q(t)$ in Fig.~\ref{fig:phisnu} for 
$\varepsilon = 0.5$ and $\varphi = 0.4$. Outside the transient regime, the 
undamped solution is shifted by 45\% compared to the damped solution. For 
the MSD in Fig.~\ref{fig:MSDnueps05} for the same values of $\varepsilon$ 
and $\varphi$, it results in a similar shift of 45\% in time scales. At 
the transition point, however, the solutions with and without damping 
deviate from each other for the relatively large window $-2<\log_{10}t<1$. 
It is also seen that the curves with Enskog damping deviate from the 
short-time asymptote $3v_0^2t^2$ considerably earlier in time. Together, 
those effects can mask the glassy dynamics expected for moderately large 
windows in time that are accessible to experiments and simulation.

\section{Summary and Outlook}

In conclusion, the results discussed above and in \cite{Kranz2010} suggest 
for testing the theory experimentally. While for the variation of the 
packing fraction, $\varphi$, one expects to recover most features known 
from the thermal glass transition --- from the variation of $\varepsilon$ 
the following scenario should emerge:

(1) The glass transition shifts to higher packing fractions $\varphi^c$ 
for lower $\varepsilon$, cf. inset of Fig.~\ref{fig:fseps}. Since the 
overall change is around 10\% it should be measurable directly. The 
behavior of time scales and diffusion coefficient of the MSD demonstrated 
in Fig.~\ref{fig:Dtau} supports the possibility of extracting this change 
from a dynamical experiment; even the asymptotic scaling law can be used 
reliably. For a discussion of experimental data in three dimensions one 
must first consider that the value for the hard sphere system in its 
elastic limit can be assumed at $\varphi^c_\text{exp} \approx 0.58$ 
\cite{Sperl2005,Goetze2009}. Hence, also the transition at smaller 
$\varepsilon$ will be shifted to yet higher values of $\varphi$.

(2) The localization length $r_c$ decreases for the mean-squared 
displacement (cf. inset of Fig.~\ref{fig:MSDcritloc}), and the 
corresponding plateau values $f^s_q$ (cf. Fig.~\ref{fig:fseps}) increase 
for smaller $\varepsilon$. While an absolute determination of $f^s_q$ and 
$r_c$ from measured $\delta r^2(t)$ and $\phi^s_q(t)$ may be ambiguous due 
to limited windows in time, a comparative measurement between different 
$\varepsilon$ should be within experimental resolution for the MSD, cf. 
Fig.~\ref{fig:MSDcritloc}.

(3) The $\alpha$-scaling function for the long-time decay of the 
correlation functions and the MSD changes with $\varepsilon$. Since 
absolute changes in the critical exponents are comparably small, see inset 
of Fig.~\ref{fig:fseps}, direct observation of changes in the exponents is 
probably rather difficult. However, a qualitative method is given by 
testing the $\alpha$-scaling: For fixed $\varepsilon$, curves for 
different $\varphi$ scale along their plateau value onto a single master 
curve, cf. Fig.~\ref{fig:phisnu} and \ref{fig:alpha}. In contrast, when 
going on a path along the transition in the $\varepsilon$ direction, a 
violation of this $\alpha$-scaling is expected.

While one can expect that the overall behavior of the MCT for granular 
systems is similar in two and three dimensions as is the case for the 
elastic hard-sphere system \cite{Bayer2007}, the necessary glass form 
factors, critical amplitudes, and exponents cannot be estimated without 
performing the actual calculation. The dynamical scenarios observed in two 
dimensions \cite{Abate2006,Gholami2011} nevertheless support the existence 
of a glass transition in a driven granular fluid. In addition, a scenario 
for the evolution of $\phi^s_q(t)$ is found in \cite{Reis2007} that is 
quite reminiscent of the 3D case shown in Fig.~\ref{fig:phisnu}. However, 
in contrast to the binary mixture in \cite{Abate2006}, the results from 
\cite{Reis2007} have to be taken with some caution as the monodisperse 
experiment has a higher tendency towards ordering which may influence the 
observed dynamics. In all two-dimensional experiments, the estimated 
plateau of the MSD is around $4r^{2}_c \approx 0.01d^2$, which is 
consistent with results anticipated from Fig.~\ref{fig:MSDcritloc}. 
However, the values reported for the localization length in 
Refs.~\cite{Menon1997a,Goldman2006} are smaller than predicted here by a 
factor of $10^3$. One possible explanation for such a deviation is the 
influence of accelerations due to gravity \cite{Durian2000} which might be 
remedied by future experiments under microgravity.

\acknowledgments%
We thank DFG for funding under FG~1394.

\end{document}